\documentclass[11pt,a4paper]{article}

\usepackage[utf8]{inputenc}
\usepackage[T1]{fontenc}
\usepackage{lmodern}
\usepackage[margin=1in]{geometry}
\usepackage{amsmath,amssymb}
\usepackage{graphicx}
\usepackage{booktabs}
\usepackage{multirow}
\usepackage{enumitem}
\usepackage{xcolor}
\usepackage{caption}
\usepackage{subcaption}
\usepackage{tabularx}
\usepackage{array}
\usepackage{float}
\usepackage{fontawesome5}
\usepackage[most]{tcolorbox}
\usepackage[colorlinks=true,linkcolor=blue!60!black,citecolor=blue!60!black,urlcolor=blue!60!black]{hyperref}
\hypersetup{
  pdftitle={Does a Toehold Make a Bidder Bolder? Preemption and Multiplicity in Multi-Round Takeover Auctions},
  pdfauthor={Zain Naboulsi},
  pdfsubject={Imperfect-information games, takeover auctions, toeholds, reinforcement learning},
  pdfkeywords={sequential auctions, toeholds, preemptive bidding, equilibrium multiplicity, imperfect-information games, policy gradient, exploitability, OpenSpiel}
}
\usepackage{natbib}

\definecolor{sparqcoral}{HTML}{C94A2A}
\definecolor{sparqcorallight}{HTML}{FFF5F2}
\definecolor{sparqcoralborder}{HTML}{E8634A}

\newcommand{\theold}{\theta}   

\begin{document}

\begin{center}
  {\LARGE\bfseries\color{sparqcoral}%
    Does a Toehold Make a Bidder Bolder?\\[0.3em]
    Preemption and Multiplicity in Multi-Round Takeover Auctions\par}
  \vspace{0.6em}
  {\large Zain Naboulsi}\\[0.3em]
  Principal AI Competency Lead, Sparq\\[0.2em]
  \texttt{zain.naboulsi@teamsparq.com}
\end{center}

\vspace{0.4em}

\begin{tcolorbox}[
  colback=sparqcorallight,
  colframe=sparqcoralborder,
  arc=3pt,
  boxrule=1pt,
  left=9pt, right=9pt, top=7pt, bottom=7pt,
  title={\textbf{Abstract}},
  fonttitle=\normalsize,
  coltitle=white,
  colbacktitle=sparqcoral,
  attach boxed title to top left={yshift=-2mm, xshift=5mm},
  boxed title style={arc=2pt, boxrule=0pt}
]
A bidder can quietly buy a stake in a company before making an offer for it. That stake, a \emph{toehold}, is supposed to pay for itself twice: it makes the bidder willing to bid harder, and it frightens rivals into staying out of the fight. The first effect is arithmetic. The second is what would justify the cost and exposure of taking a toehold at all, and it is why the rarity of toeholds in practice has been a standing puzzle \citep{bettoneckbothorburn2009}.

We ask whether the second effect is there once the contest is modeled as several rounds of escalating bids rather than the single exchange classical models assume. We cast a multi-round ascending common-value takeover auction with toeholds and jump bidding as a two-player extensive-form game of imperfect information, and solve it for own-profit Bayes-Nash equilibria certified as $\varepsilon$-equilibria with $\varepsilon$ between $5\times10^{-7}$ and $8\times10^{-5}$. Three things follow. The auction pins down what the toehold-holder earns but not how it behaves. This \emph{equilibrium multiplicity} is not numerical slack: profiles certified at $\varepsilon\sim10^{-6}$ deter the rival with probability $0.000$ and $0.333$, a gap in conduct five orders of magnitude larger than the $\varepsilon$ that certifies them, and an exact best-response calculation rather than solver output confirms each conduct is a best reply to the other. Preemptive jump bidding persists at a \emph{zero} toehold, recovering Fishman-style signalling inside a solved game and showing the stake buys none of it. And in the instance we solve, the classical link from toehold size to deterrence holds only at one round; a genuine second round flattens it.

So the two reasons to buy a toehold do not survive equally. The profit reason does; the deterrence reason is not identified, which offers a model-side reason toeholds may be rarer than theory expects, alongside the procedural explanations, disclosure thresholds and price impact, that our model does not represent. One warning generalizes past takeovers: solve this game once and it returns a confident figure for what a preemptive bid is worth, solve it again from a different start and it returns a different one, equally converged, on the same auction. We also benchmark solvers on it, where a generic policy-gradient method reaches near-exact equilibrium and NFSP fails to converge, and we extend past the size exact methods can enumerate. All games, solvers, and experiments are released.

\vspace{0.5em}
\noindent\begin{minipage}[b]{0.82\linewidth}
  \footnotesize\raggedright\mbox{\faGithub\ \url{https://github.com/zainnab-sparq/sequential-takeover-auctions}}
\end{minipage}%
\hfill
\begin{minipage}[b]{0.16\linewidth}
  \raggedleft
  \includegraphics[height=1cm]{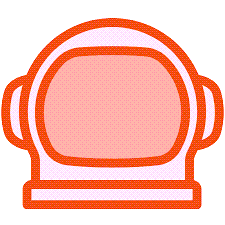}\\[-2pt]
  {\small\textsf{\textbf{Sparq}}}
\end{minipage}
\end{tcolorbox}

\vspace{0.4em}
\noindent\textbf{Keywords:} sequential takeover auctions, toeholds, preemptive bidding, equilibrium multiplicity, imperfect-information games, policy gradient, exploitability, OpenSpiel

\section{Introduction}

Before making a takeover bid, an acquirer can quietly buy a block of the target's shares on the open market. That block is a \emph{toehold}, and whether to take one is a live decision with real money behind it. A toehold must be disclosed once it crosses a regulatory threshold, the buying itself can tip the market and raise the price the acquirer eventually pays, and it can make an approach that was meant to be friendly look hostile. Deal teams accept those costs when they believe the toehold buys something worth more. What the textbook says it buys is a bidding edge. This paper asks whether that edge is there once the contest runs over several rounds of escalation rather than the one or two moves the classical models allow, and finds that it is not. We reproduce the classical prediction at one round, at the two positive toeholds we sweep there, and then watch it disappear as the contest lengthens.

The textbook argument is arithmetic. A bidder holding a toehold comes out ahead either way. If it wins, it pays the full price only for the shares it does not already own. If it loses, it sells its block into the winner's price, and the higher that price goes the more it collects. Both branches are better than they would be with no toehold, so the holder can afford to bid harder, and a rival who works that out may decide not to compete at all. That is the toehold's celebrated deterrence or ``aggressiveness'' channel \citep{bulow1999,fishman1988}. Throughout, \emph{deterrence} means the chance that the rival, having seen the opening bid, concedes on the spot rather than fight for the company. It is also the source of a long-standing empirical puzzle: if a toehold confers a bidding edge, bidders should acquire toeholds routinely before launching a bid, yet \citet{bettoneckbothorburn2009} document that they seldom do. Explanations invoke the cost of tipping off the market, legal disclosure thresholds, or the risk of being seen as hostile; \citet{daigryglewiczsmit2021} argue instead that toeholds are effective once the selection into taking one is corrected for, so the puzzle is partly an artifact of who chooses to take one. We ask a different question: is the predicted bidding edge even there, once the contest is modeled as it actually plays out, over several rounds of escalation rather than a single sealed bid or a two-move exchange?

\paragraph{Why this matters.} The deterrence argument is one of the two reasons anyone gives for buying a toehold, and it is the one that is supposed to change how a rival behaves. Our answer is that it does not survive once the contest is given more than one or two moves. The toehold still raises what its holder earns, so the profit argument stands; what does not stand is the claim that owning a stake reliably frightens a rival out of the auction. A deal team weighing the disclosure and price-impact costs of a toehold against an expected deterrence benefit is, on the evidence here, weighing them against something the model cannot deliver. For auction theory the point is narrower and sharper: the tidy ``bigger toehold, more deterrence'' relationship is a property of our model when it is stopped after one or two moves, and does not survive in that same model once the contest has a genuine second round. And for anyone computing economics out of a game solver, the paper is a worked warning. Run the solver once and it hands back a confident number for how much preemption is worth. Run it again from a different starting point and it hands back a different one, equally converged, on the same game. The number was never a property of the auction; it was a property of where the solver happened to land.

Our companion paper \citep{naboulsi2026diligence} studied \emph{sealed}, simultaneous takeover auctions, where each bidder submits one bid and the toehold's aggressiveness channel is competed away, so a sealed model cannot express preemption at all. That paper asserted, but could not show, that the channel ``lives in sequential, asymmetric contests.'' This paper makes the sequential contest the object of study and settles the assertion in two parts, confirming the structural half and contradicting the toehold half. We model a multi-round ascending auction in which a toehold-holder and a rival take turns, each able to jump to any higher price on a fixed ladder of prices or to pass. The bid history is public, so each side sees what the other has bid; each side's due-diligence findings are private. The target has a \emph{common value}: it is worth the same to both bidders, and neither knows what that is, so each must infer it partly from the other's bidding. We then solve the game, meaning we compute strategies from which neither bidder can profitably deviate given what it knows, and we score each solution by how much a bidder could gain by deviating. That number, $\varepsilon$, is the paper's unit of trust: an $\varepsilon$ of $10^{-6}$ means no bidder can improve its profit by more than a millionth by playing anything else, so no bidder has much reason to move, and a referee can check the claim by re-running it. We state every claim in those terms rather than in terms of exact equilibria, because in a flat game a profile with small $\varepsilon$ can still sit some distance from an exact one. Crucially, we also measure what a preemptive jump bid is worth using a construction that does not depend on the solver landing on any particular solution.

The results overturn the simple story in three steps. First, the game has more than one solution, and they disagree about behavior. Started from different points, the solver lands on the same profit for the toehold-holder (to four decimals at two of the three toeholds we certify, and to within $6.3\times10^{-5}$ at the third) but on completely different conduct: in one solution the holder jump-bids and the rival folds, in another it opens cheaply and nobody folds. Both are stable, because both are circular in a way that holds together. ``I jump because you will fold, and you fold because I jumped'' is self-consistent. So is ``I open low because you will not fold, and you do not fold because I opened low.'' Nothing in the auction picks between them. Second, the jump-bidding solution still exists when the toehold is \emph{zero}, so preemption is a feature of taking turns in public, not something the toehold purchases. Third, the tidy ``bigger toehold, more deterrence'' relationship that classical two-move models deliver turns out to be an artifact of stopping after two moves. It holds at one round, stops responding at two, and by three rounds the solutions multiply and deterrence is no longer pinned down at any toehold. The channel does not fade gradually as the contest lengthens; it is gone by the second round.

We make three contributions. First, to our knowledge this is the first time a genuinely multi-round takeover auction with a toehold has been solved computationally to a checkable standard of accuracy. That includes a tool we build for the purpose: instead of asking the solver what the holder does, we force it to open at each price in turn, let every later decision play out optimally, and read off what each opening earns. That calculation is exact given the rival it is run against, so it contributes no convergence error of its own. What it exposes rather than removes is the dependence on \emph{which} rival, which is why we report an interval instead of a number. Second, we establish the two economic findings above, that the solutions are multiple and that the toehold-to-deterrence channel is an artifact of short models, and reconcile them with the existing signalling explanation of jump bidding \citep{fishman1988} and with the toehold puzzle. Third, we test which solvers can actually handle this game, including versions of it too large to write down completely, and report which methods still work there. All games, solvers, and experiments are released.

\paragraph{Who should read what.} A reader who wants the economics can take Sections~\ref{sec:model} through~\ref{sec:reconcile} together with the limitations in Section~\ref{sec:limitations}, which bound every claim in them; the two tables and the one figure carry the argument. A reader who wants to know whether to believe it should read the methodological caution in Section~\ref{sec:multiplicity} and the limitations in Section~\ref{sec:limitations}, which is where we say what the results do not show. Sections~\ref{sec:benchmark} and~\ref{sec:intractable} are about solvers rather than auctions and can be read independently of both.

\section{Related work}

\paragraph{Toeholds and preemptive bidding.} The idea that a toehold makes a bidder more aggressive, lowering a rival's chance of winning, is classic \citep{bulow1999,klemperer1999}. Preemptive \emph{jump} bidding, bidding above the minimum required to signal strength and deter entry, was formalized by \citet{fishman1988} as a \emph{signalling} phenomenon: a high bid credibly conveys a high valuation, and a rival holding a weak signal infers that competing means overpaying, so it withdraws. Crucially, Fishman's mechanism requires no toehold. \citet{avery1998} showed jump bidding can arise even in a setting with no private information to signal, as a coordination device. \citet{danielhirshleifer2018} develop a costly sequential bidding model in which each bid entails a cost, generating preemptive first bids and infrequent competition. Closest to our setting, \citet{dodonova2012} studies toeholds and signalling in takeover auctions in a compact model; our multi-round treatment reproduces the classical comparative static, that a larger toehold deters more, in a one-round contest, and then shows it does not survive additional rounds. Empirically, toeholds are rarer than the classical edge would predict (the toehold puzzle), a tension our multiplicity and two-move-artifact findings speak to directly.

\paragraph{Common-value auctions and the winner's curse.} Bidding in a common-value auction requires shading against the adverse selection of winning \citep{capen1971,milgromweber1982,wilson1977}. Our game is common-value with noisy private signals, and the preemption mechanism we find is a winner's-curse inference: a rival that reads a jump bid as evidence of a high common value folds, because competing would mean winning mainly in the states where its own signal is misleadingly high. This is the same inference as in \citet{fishman1988}, realized computationally.

\paragraph{Computational game solving.} Exact and sampling solvers for imperfect-information games, counterfactual regret minimization \citep{cfr2007,deepcfr2019}, magnetic mirror descent \citep{mmd2023}, and policy-space response oracles \citep{psro2017}, have mostly been developed on recreational benchmarks. Generic deep reinforcement learning, PPO \citep{ppo2017}, PPG \citep{ppg2021}, and NFSP \citep{nfsp2016}, has been argued to rival specialized machinery on such games \citep{rudolph2026}, and multi-agent reinforcement learning is increasingly used to analyze auction mechanisms directly \citep{deon2024}. Sequential, asymmetric games are the setting where specialized regret-based solvers are expected to have the largest edge, so they are a sharp test. We benchmark both families on our sequential auction and, following our companion paper, use a learned-best-response \emph{exploitability} estimate \citep{openspiel2019} when the game grows beyond exact enumeration. Exploitability is the standard score for a strategy in these games: freeze it, then measure how much a perfect opponent could take from it, so zero means an exact equilibrium and lower is better.

\section{A multi-round takeover auction with a toehold}
\label{sec:model}

\paragraph{The contest in words.} Two firms want the same target. The target is worth some amount that neither of them knows, and it is worth the same to both, so the problem is not disagreement about taste but ignorance about a fact. Each firm does its own due diligence and comes away with a noisy read on that number. They then bid against each other in turns, each seeing every bid the other makes but never seeing the other's homework. On its turn a firm can raise to any higher price it likes, not merely the next one up, which is what makes an aggressive opening possible in the first place, or it can walk away. One of the two firms already owns a slice of the target. Everything that follows is the formal version of that paragraph.

\paragraph{Primitives.} Two bidders compete for a target of common value $W$, drawn uniformly from a grid of $\mathrm{num\_values}$ levels. Both are risk-neutral, meaning each one values a gamble at its average payoff and so has no separate appetite for or aversion to risk. Each bidder $i$ receives $k=\mathrm{num\_signals}$ noisy private signals of $W$; with probability $1-\mathrm{signal\_noise}$ a signal equals $W$ and otherwise is uniform on the value grid. Bidder $0$ holds a toehold $\theold\in[0,1)$: a fraction of the target it already owns. Prices lie on a discrete grid of $\mathrm{num\_bids}$ levels spaced by $\mathrm{bid\_step}$.

\paragraph{Protocol.} The auction is an ascending, alternating-move contest of up to $R=\mathrm{num\_rounds}$ rounds. A round is one turn for each bidder, so an $R$-round contest is $2R$ moves and $R=1$ is the two-move exchange, one bid and one response, that the classical models analyze. We count in rounds throughout. On its turn a bidder either \emph{raises}, jumping to any grid price strictly above the standing bid (a jump bid is thus available, not just the minimum increment), or \emph{passes}. Passing while behind ends the auction and the standing bidder wins; both passing with no standing bid ends it with no sale; otherwise play continues until the round limit. The bid history (the standing bid, whose turn it is, and the round index) is public and carried in each bidder's information state; a bidder's own signals stay private; the game has perfect recall. Because raises must strictly increase, the number of raises is bounded by the price grid: the tree depth is capped by $\mathrm{num\_bids}$, not by $R$, so added rounds deepen the contest only while $\mathrm{num\_bids}\gtrsim 2R$.

\paragraph{Payoffs.} If bidder $1$ wins at price $p$ it earns $W-p$ and the toehold-holder earns $\theold\, p$ (it sells its share into the winning price). If the toehold-holder wins at price $p$ it earns $W-(1-\theold)p$, paying only for the fraction it does not already own. The setup matches the payoff contract of our companion sealed game, so the two are directly comparable on an identical grid. Note that $R=1$ is \emph{not} that sealed game: the rival still moves second and still sees the opening bid, which is exactly why deterrence is a meaningful quantity at $R=1$ and is not one in a sealed auction.

\paragraph{The instance.} All the economics below is one game: $\mathrm{num\_values}=3$, $\mathrm{num\_bids}=9$ with $\mathrm{bid\_step}=0.375$ (so the grid spans $[0,3]$, the full range of $W$), $k=1$ signal per bidder, $\mathrm{signal\_noise}=0.5$ for both bidders, and $R$ swept over $\{1,2,3\}$. Each cell is solved from $6$ restarts (restart $0$ uniform, the rest drawn from seed $20{,}260{,}713$). The certified runs use a $2\times10^{6}$-iteration budget with the tolerance set to $10^{-8}$, and a restart enters a table only if its own-profit NashConv is at or below $10^{-4}$; both the budget and that acceptance bound matter, and both are recorded per row in the released data. The benchmark, scaling and intractable-regime instances of Sections~\ref{sec:benchmark} and~\ref{sec:intractable} are different and smaller, and are stated there.

\paragraph{Encoding.} The game is implemented as an OpenSpiel \citep{openspiel2019} extensive-form game with the bid history as public observation tokens and the private signals as one-hot features. Perfect recall and the information-set structure (own signals hidden, bid history public) are enforced by construction and checked by unit tests; exact counterfactual regret minimization drives exploitability below $0.01$ on small instances, confirming the solver converges on the encoding as expected.

\section{Solving for own-profit equilibria}

Most off-the-shelf solvers assume the game is \emph{zero-sum}: whatever one side gains, the other loses. A takeover auction is not like that. Both bidders can do badly by overpaying, and both can do well; what each one cares about is its own profit, not the difference between them. Games of that kind are called \emph{general-sum}, and the distinction is not cosmetic here, because a solver aimed at the zero-sum version would be maximizing a rivalry the auction does not contain.

So we solve for the profit each bidder actually earns. The method is \emph{fictitious play}, which is easier than its name: each side repeatedly works out the best reply to the average of everything the other side has done so far, and those replies are themselves averaged over time. Working out a best reply here means walking the game tree and scoring each ending by the profit it pays, rather than by a win or a loss. After each round we measure how much either bidder could still gain by abandoning the current strategies and best-responding instead. That quantity is the $\varepsilon$ of the introduction, and it is the certificate: when it is small, the strategies are close to an equilibrium, and how small it is says how close. Two properties of this problem shape everything that follows. Finding an equilibrium in a general-sum game is PPAD-hard, meaning no efficient general method for it is known, and such a game may have many equilibria rather than one. So nothing entitles us to a unique answer, and we treat convergence as something to be measured and reported rather than assumed.

\paragraph{Reading economics without trusting the solver.} Reading a result off one solver run is unreliable, because the run quietly picks one equilibrium out of however many exist and then reports its properties as though they were the game's. We therefore add a second instrument that does not depend on the solver at all. We force the toehold-holder to open at a chosen price, let every decision after that play out optimally, and record what it earns. Repeating this for each price on the ladder traces out what every possible opening bid is worth, including the aggressive ones. Two things make this trustworthy where a single solve is not. It is an exact calculation rather than an iterative approximation, so there is no convergence to doubt, and we check it against brute force over every possible opening. The result is a way to price a preemptive jump bid that owes nothing to fictitious play having settled \emph{for the bidder whose opening we force}, which matters precisely because whether it settles is the question at issue. The rival it is measured against is still solver output, and that is where the remaining ambiguity lives.

\section{Equilibrium multiplicity: same value, different deterrence}
\label{sec:multiplicity}

The central finding is that the auction pins down what the toehold-holder \emph{earns} but not how it \emph{behaves}. Two solutions can pay the holder the same amount to four decimal places and yet look nothing alike at the table: in one it opens with an aggressive jump and the rival concedes a third of the time, in the other it opens cheaply and the rival almost never concedes. That concession rate is the deterrence of the introduction, and it is measured here as the probability mass on the rival conceding immediately after the opening bid. It is the quantity the whole toehold literature is about, and it is the one this auction refuses to fix. Independent solver restarts converge to the same own profit for the toehold-holder, to four decimals at $\theold=0$ and $\theold=0.15$ and to within $6.3\times10^{-5}$ at $\theold=0.05$, and to completely different conduct. Table~\ref{tab:multiplicity} reports representative pairs. At a toehold of $0.05$, two profiles worth $0.2716$ and $0.2715$ to the toehold-holder deter the rival with probability $0.002$ and $0.333$ respectively; the preempting profile is certified as an $\varepsilon$-Nash equilibrium with $\varepsilon\approx 4\times10^{-6}$ and the passive profile, which converges more slowly, with $\varepsilon\approx 8\times10^{-5}$.

The contrast is not uniform across the toehold range, and where it fades it does so for two different reasons. It is sharpest at $\theold=0$ ($0.000$ versus $0.333$) and narrows to $0.220$ versus $0.333$ by $\theold=0.15$. At $\theold=0.20$ every accepted restart lands on the same conduct ($0.3333$ to four decimals), so there is nothing to certify. Above $\theold=0.20$ no restart converges in the base three-round game at the budget of the nine-point sweep, so deterrence is \emph{unidentified} there, meaning the model does not single out one value for it, by non-convergence rather than by multiplicity: four of those nine toeholds ($0.25$, $0.30$, $0.40$, $0.50$) yield no usable equilibrium and are absent from the released summary.

\begin{table}[H]
\centering
\small
\begin{tabular}{ccccc}
\toprule
Toehold $\theold$ & \multicolumn{2}{c}{Bidder-0 value} & \multicolumn{2}{c}{Rival deterrence} \\
\cmidrule(lr){2-3}\cmidrule(lr){4-5}
 & profile A & profile B & profile A & profile B \\
\midrule
$0.00$ & $0.1840$ & $0.1840$ & $0.000$ & $0.333$ \\
$0.05$ & $0.2716$ & $0.2715$ & $0.002$ & $0.333$ \\
$0.15$ & $0.4554$ & $0.4554$ & $0.220$ & $0.333$ \\
\bottomrule
\end{tabular}
\caption{Equilibrium multiplicity in the three-round auction. Each row shows two profiles, each an $\varepsilon$-Nash equilibrium, that give the toehold-holder the same value but deter the rival with very different probability. Certified $\varepsilon$ by row: $2.4\times10^{-6}$ at $\theold=0$, $3.5\times10^{-6}$ at $\theold=0.15$, and $7.6\times10^{-5}$ at $\theold=0.05$, where the passive profile is the slower of the pair (its partner is at $4.3\times10^{-6}$). Values agree to four decimals except at $\theold=0.05$, where they differ by $6.3\times10^{-5}$, less than that row's $\varepsilon$, so the two values are not distinguishable at the precision we certify. Deterrence is not identified by the toehold. Regenerate with \texttt{sequential\_general\_sum.py certify}, or in one second from the committed per-restart data with \texttt{certificate\_from\_selection\_csv}.}
\label{tab:multiplicity}
\end{table}

\begin{figure}[H]
\centering
\includegraphics[width=\textwidth]{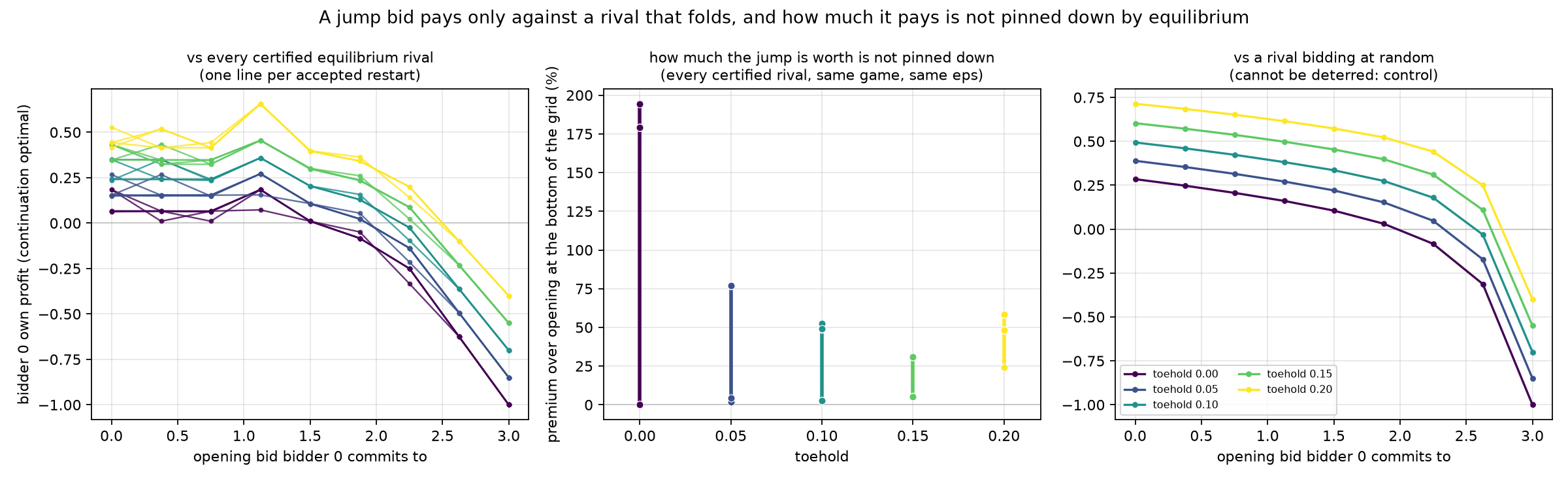}
\caption{Forced-opening best-response profit for the toehold-holder as a function of the opening bid it commits to, with every later decision best-responding. Left: one curve per certified equilibrium rival, colored by toehold. Curves that peak at the preemptive jump at $1.125$ and curves that peak at the cheapest opening both occur at the same toehold, which is the multiplicity. Middle: the resulting jump premium, one point per certified rival with the bar spanning them. The spread is the finding. Rivals certified at the same $\varepsilon$, several of them deterring with the same probability to four decimals, price the jump an order of magnitude apart, so no single number is the premium and any figure that showed one would be showing its own selection rule. Right, the control: against a rival bidding at random, which cannot be deterred, profit slopes straight down and the premium is exactly zero at every toehold. A jump bid only pays when it can scare a rival off, which is why both the preempting and the non-preempting conduct are self-consistent equilibria. The forced-opening best response is exact and the control panel is solver-free; the equilibrium rivals are solver output, budget-terminated at the same probe settings as Tables~\ref{tab:multiplicity} and~\ref{tab:rounds}, at NashConv from $2\times10^{-6}$ to $7.6\times10^{-5}$, the looser end being the rivals that carry the low-premium end of the spread.}
\label{fig:mechanism}
\end{figure}

\paragraph{Why, mechanically.} The forced-opening profit curve shows both conducts are self-consistent (Figure~\ref{fig:mechanism}). The cleanest case is the one that carries the second finding. At $\theold=0$ the six certified restarts split in two: four deter the rival with probability $0.333$, and two essentially never do ($7\times10^{-7}$). Against the four, committing to the jump at $1.125$ rather than opening at the bottom of the grid raises the toehold-holder's profit by $179$ to $194\%$; against the two it raises it by exactly nothing, and the curve peaks at the cheapest opening. So ``I preempt because you fold, you fold because I preempt'' and ``I open low because you will not fold, you will not fold because I open low'' are both equilibria. This is coordination-style multiplicity, not solver failure.

\paragraph{Is it just $\varepsilon$ slack?} The obvious objection to any computational multiplicity claim is that two profiles certified only to $\varepsilon$ need not be two equilibria. Where payoffs are nearly indifferent, or at information sets the profile rarely reaches, strategies are close to unconstrained, and a difference that looks behavioural could be numerical room. Two things answer it here. First, the scale is wrong for slack: at $\theold=0$ the two conducts deter with probability $0.000$ and $0.333$ while both are certified at $\varepsilon\approx2.4\times10^{-6}$, so the gap in behaviour is about $10^{5}$ times the gain either bidder could get by deviating. Second, and more decisively, the forced-opening curve settles it without the solver. It is an exact calculation, validated against brute-force enumeration over every pure opening policy, and it shows each conduct is an exact best reply to its counterpart rather than a near-tie: against the deterring rivals, committing to the jump earns $179$ to $194\%$ more than opening at the bottom of the grid, and against the passive ones it earns nothing at all and the curve peaks at the cheapest opening. A profile cannot be slack in a direction that costs a bidder nothing and gains it nothing. What we do not claim is an enumeration of the equilibrium set: we exhibit two conducts that are each supported as best replies, not a proof that these are the only ones. That gap is not one more compute could close. The tree of the three-round instance has about $40{,}000$ nodes, which is small, but the bidders hold $768$ and $1{,}146$ information sets between them, so the space of pure strategy profiles is on the order of $10^{537}$. A small game tree does not imply a small strategy space, and that is precisely why the equilibrium set has to be probed by restarts and certified pointwise rather than enumerated.

There is a sharper version of this, and it is worth stating separately. Not only is the deterrence rate unpinned, so is the amount of money a jump bid is worth. Non-identification bites harder here than the phrase suggests: several answers are equally consistent with everything the model says, so quoting one of them is a choice rather than a finding. The deterrence rate is a single number standing in for a whole strategy, one that specifies what the rival would do in many situations, most of which never arise in the solution being summarized. The forced-opening curve puts a price on exactly those unvisited situations. That is why two solutions can report the same deterrence rate and still disagree about what preemption earns, and they do: at $\theold=0.15$ four certified profiles deter the rival with probability $0.3333$ to four decimals and yield premia of $31\%$, $31\%$, $5\%$ and $5\%$, and at $\theold=0.20$ four profiles that all deter at $0.3333$ yield premia from $24\%$ to $58\%$. Reporting a single premium per toehold would therefore be reporting a choice of representative profile rather than a property of the game. We report the interval spanned by the certified rivals instead: $0$ or $179$ to $194\%$ at $\theold=0$, $2$ to $77\%$ at $\theold=0.05$, $3$ to $52\%$ at $\theold=0.10$, $5$ to $31\%$ at $\theold=0.15$, and $24$ to $58\%$ at $\theold=0.20$. These are lower bounds on the true spread, since they span the equilibria that six restarts happen to find rather than the equilibrium set.

As a control, against a \emph{uniform}, unbluffable rival the forced-opening curve slopes straight down: the best opening is the cheapest one at every toehold and the premium is exactly zero. A jump bid buys deterrence, and deterrence is worthless against a rival that cannot be deterred. That panel is the one exhibit here with no solver in it, and it is what licenses reading the premium as the price of deterrence rather than as an artifact of the forced-opening construction. Premia are measured against the profit of opening at the bottom of the grid; the released data reports that normalization and the peak-relative one in separate columns, one row per certified rival.

\paragraph{Preemption at a zero toehold.} The decisive observation is in the first row of Table~\ref{tab:multiplicity}. At $\theold=0$, some restarts still converge on a profile in which the toehold-holder jump-bids and the rival folds a third of the time, certified at $\varepsilon\approx 2.4\times10^{-6}$. \emph{Preemption is not something the toehold buys.} The toehold sets the toehold-holder's profit; the sequential structure is what makes preemption possible; which equilibrium is played is a matter of coordination. The original hypothesis, that a toehold buys a preemptive jump bid and revives the aggressiveness channel, is false as a causal claim.

\paragraph{A methodological caution.} Reaching this conclusion required checking the full set of restarts, not a sample. An early analysis re-ran three restarts at a tight tolerance, saw them agree, and wrongly concluded a competing equilibrium was a numerical artifact; two other restarts converged to it, and the released rows for them certify at $\varepsilon$ of $2.1\times10^{-6}$ and $2.2\times10^{-6}$. When testing for multiplicity, a sample of restarts that agree proves nothing, because the disagreeing basin may simply not be in the sample. Likewise, a fictitious-play run that stops \emph{at} its tolerance has not converged; it has been let go. Every number in Tables~\ref{tab:multiplicity} and~\ref{tab:rounds}, and every rival policy behind Figure~\ref{fig:mechanism}, was re-run with the tolerance set far below what the run can reach, so the compute budget, not the stopping rule, ended it. One further disclosure belongs here, since it is the same concern one level down. The $\varepsilon$ we report is the tightest profile the run visited, not the one it happened to stop on. This is sound, because a profile whose measured NashConv is $\varepsilon$ \emph{is} an $\varepsilon$-equilibrium whenever it was reached, but it is not the same as the run having settled there, and own-profit fictitious play in a general-sum game can orbit rather than converge. The released data therefore carries a \texttt{wander} column, the ratio of a solve's final $\varepsilon$ to its certified one, so a reader can tell the two apart: $1.0$ means the run settled and a large value means it cycled and the certificate caught a favourable moment. The cells behind Tables~\ref{tab:multiplicity} and~\ref{tab:rounds} are clean on this measure. Some cells in the grid-refinement study are not, and we flag them where we use them. A related trap sits one level up, in how an exhibit chooses which certified profile to show. The mechanism figure originally picked its rivals by extreme deterrence and the certificate picked them by tightest $\varepsilon$ within a tie band, so the two exhibits could represent the same toehold by different profiles; at $\theold=0.15$ they did, and the premium the figure reported moved by a factor of six depending on which rule was in force. Where a quantity varies across equally certified profiles, selecting one is an editorial act, and the honest response is to report the spread rather than to standardize the selection rule.

\paragraph{Does it survive grid refinement?} The multiplicity is a statement about a discretized price grid, so the obvious worry is that it is an artifact of the discretization. We solve the same price range at three resolutions, all at the certified budget: seven levels (step $0.500$), the base nine (step $0.375$), and thirteen (step $0.250$), six restarts at each of three toeholds, fifty-four solves. The answer is that refinement does not destroy the multiplicity; coarsening does. At thirteen levels the two certified profiles at $\theold=0$ deter with probability $0.005$ and $0.333$, the same qualitative split as at nine levels, and both settled rather than cycling, in the sense defined above: their final and certified $\varepsilon$ agree. At seven levels, by contrast, the low-deterrence conduct never appears at all: the minimum deterrence over every accepted solve is $0.333$ at $\theold=0$ and $0.393$ at $\theold=0.30$. Both minima come from solves that settled rather than cycled, which matters because one other accepted solve at that resolution did cycle badly (its final $\varepsilon$ is $328$ times its certified one), so the claim does not rest on it. So the coarse grid is the one that cannot express the multiplicity, and the deterrence band does not close as the grid refines, which is what this study was built to ask.

Two things do degrade with refinement, and both are worth stating. Acceptance is not monotone in tree size: eleven of eighteen solves clear $10^{-4}$ at nine levels, but only six of eighteen at seven and two of eighteen at thirteen, so the coarsest grid is not the easiest to solve. And at thirteen levels the \emph{value} half of the finding fails. The two certified profiles there are worth $0.1736$ and $0.2078$ to the toehold-holder, a gap of $0.034$ against $\varepsilon$ of $1.3\times10^{-5}$ and $8.2\times10^{-5}$, so they are distinguishable by several hundred multiples of their own certification. Value-pinning is a per-configuration observation throughout this paper rather than a theorem, and this is a third configuration where it does not hold, alongside $R=1$ and $R=2$ at a zero toehold. Deterrence non-identification, which is the claim the paper rests on, survives both resolutions that exhibit the passive branch, and even the coarse grid fails to pin deterrence down, where certified deterrence still spreads from $0.333$ to $0.408$ at $\theold=0$.

\section{The two-move artifact}

Sweeping the number of rounds turns the project around. Table~\ref{tab:rounds} reports deterrence as a function of the toehold at one, two, and three rounds. Every cell with a number in it is certified at $\varepsilon$ on the order of $10^{-6}$ to $10^{-5}$ (verified with the tolerance set to $10^{-8}$ and the run ended by budget, far tighter than the default stopping rule); the one cell without a number is the one where no restart reaches the acceptance bound at all, and it is reported as such.

\begin{table}[H]
\centering
\small
\begin{tabular}{cccc}
\toprule
& \multicolumn{3}{c}{Rival deterrence probability} \\
\cmidrule(lr){2-4}
Toehold $\theold$ & $R=1$ round & $R=2$ rounds & $R=3$ rounds \\
\midrule
$0.15$ & $0.333$ (unique) & $0.333$ (saturated) & $0.220$ or $0.333$ (multiple) \\
$0.30$ & $0.583$ (unique) & $0.333$ (saturated) & unconverged \\
\bottomrule
\end{tabular}
\caption{The toehold-to-deterrence channel by number of rounds. At one round the equilibrium is unique (at a positive toehold) and deterrence rises with the toehold, exactly as classical two-move intuition predicts. At two rounds deterrence saturates and stops responding to the toehold. By three rounds the equilibria multiply and deterrence is no longer pinned by the toehold: it spans $0.220$ to $0.333$ at $\theold=0.15$, and $0.000$ to $0.333$ at a zero toehold (a row not shown here), while at the largest toehold no restart converges even at the tightest budget.}
\label{tab:rounds}
\end{table}

At \emph{one} round, and a positive toehold, the equilibrium is unique: independent restarts land on identical numbers, and deterrence is markedly higher at the larger toehold ($0.333$ at $\theold=0.15$, $0.583$ at $\theold=0.30$, with all six restarts identical at each). We sweep only these two positive toeholds at $R=1$, so this is an increase between two points rather than a monotone curve. This reproduces the classical comparative static, in our model, at one round. At \emph{two} rounds deterrence \emph{saturates}: raising the toehold from $0.15$ to $0.30$ moves deterrence not at all ($0.333$ in both cases), whereas a two-move contest bought a large jump ($0.333\to0.583$). By \emph{three} rounds the equilibria multiply, deterrence spans $0.000$ to $0.333$ across profiles and toeholds (and $0.220$ to $0.333$ at $\theold=0.15$ alone), and preemption is available even at a zero toehold; at the largest toehold ($\theold=0.30$) no restart converges at all, even at the tightest budget, so deterrence is not identified there by any route.

So the toehold's deterrence channel behaves exactly as auction theory says \emph{in a two-move contest} and \emph{dissolves} in a genuinely multi-round one. The founding qualifier of this project, that a toehold's value compounds across three or more rounds, had the right axis and the wrong sign: the effect does not compound, it evaporates. One caveat must be stated: at a \emph{zero} toehold the two-move game already carries several equilibria, five accepted restarts spanning three distinct values, so multiplicity is not created by adding rounds in the abstract; it is created by adding rounds \emph{at a positive toehold}, which is the case of interest.

\section{Reconciliation and the toehold puzzle}
\label{sec:reconcile}

These findings do not contradict the literature; they sharpen it. For \citet{fishman1988}, preemptive jump bidding is a \emph{signalling} phenomenon, not a toehold phenomenon, and the mechanism we find is the same winner's-curse inference: the toehold-holder jump-bids, and a rival holding a weak signal infers that competing means paying at least the jump and winning mainly where its own signal is wrong, so it folds. No toehold is required for that, which is exactly why preemption appears at $\theold=0$. What dies is only the narrower, Bulow-Huang-Klemperer-flavored claim that a \emph{toehold} adds a deterrence channel on top of the signalling one. We therefore reproduce Fishman-style signalling preemption computationally and show the toehold does not add to it.

This speaks to the toehold puzzle. If a toehold buys profit but no identified deterrence, then part of the rarity that \citet{bettoneckbothorburn2009} document may need no appeal to hidden costs: in our model the theoretical bidding edge that toeholds are supposed to confer is not identified once the contest runs past one round. This is a model-side reason the edge may be weaker than assumed, complementary to the cost-based explanations rather than a replacement for them, and it does not speak to the selection correction of \citet{daigryglewiczsmit2021}, which is an empirical matter our computation cannot settle. The aggressive-toehold equilibrium requires the rival to \emph{read} a jump bid as strength; absent that coordination, the toehold still raises the holder's profit but buys no deterrence.

\paragraph{What may and may not be claimed.} We can say that preemption is \emph{sustainable} in the sequential contest and \emph{absent} in the sealed auction of our companion paper (whose equilibrium bid is flat in the toehold on the identical grid). What that demonstrates is the structural half of the companion paper's assertion: a sequential contest can express preemption where a sealed one cannot. It does \emph{not} vindicate the \emph{toehold} half of that assertion, which our zero-toehold result contradicts. We can also say that the toehold-holder's equilibrium profit rises robustly with the toehold. We do \emph{not} claim that a toehold implies any particular amount of deterrence, or any smooth deterrence-versus-toehold curve. The deterrence prediction is not identified, and drawing that curve from a single family of solves would be precisely the plausible-but-wrong result our methodology is built to avoid.

\section{Benchmarking solvers on the sequential game}
\label{sec:benchmark}

The rest of the paper is about auctions; this section and the next are about the tools. The question is a practical one for anyone who wants to run an analysis like ours: which methods actually solve a game of this shape, on an ordinary CPU, with no spending on frontier models?

Two families compete. The specialists, of which counterfactual regret minimization (CFR) is the best known, exploit the structure of imperfect-information games and come with convergence guarantees, but they need to enumerate the game, which becomes impossible as it grows. The generalists are reinforcement-learning methods borrowed from elsewhere, principally PPO and its variant PPG, which learn by playing against themselves and never enumerate anything, so they keep running when the specialists cannot. NFSP is the standard imperfect-information baseline from the RL side and serves as the comparison point within that family. We score every method by exploitability, as defined above: freeze what it learned, then compute how much a perfect opponent could take from it.

\begin{figure}[H]
\centering
\includegraphics[width=\textwidth]{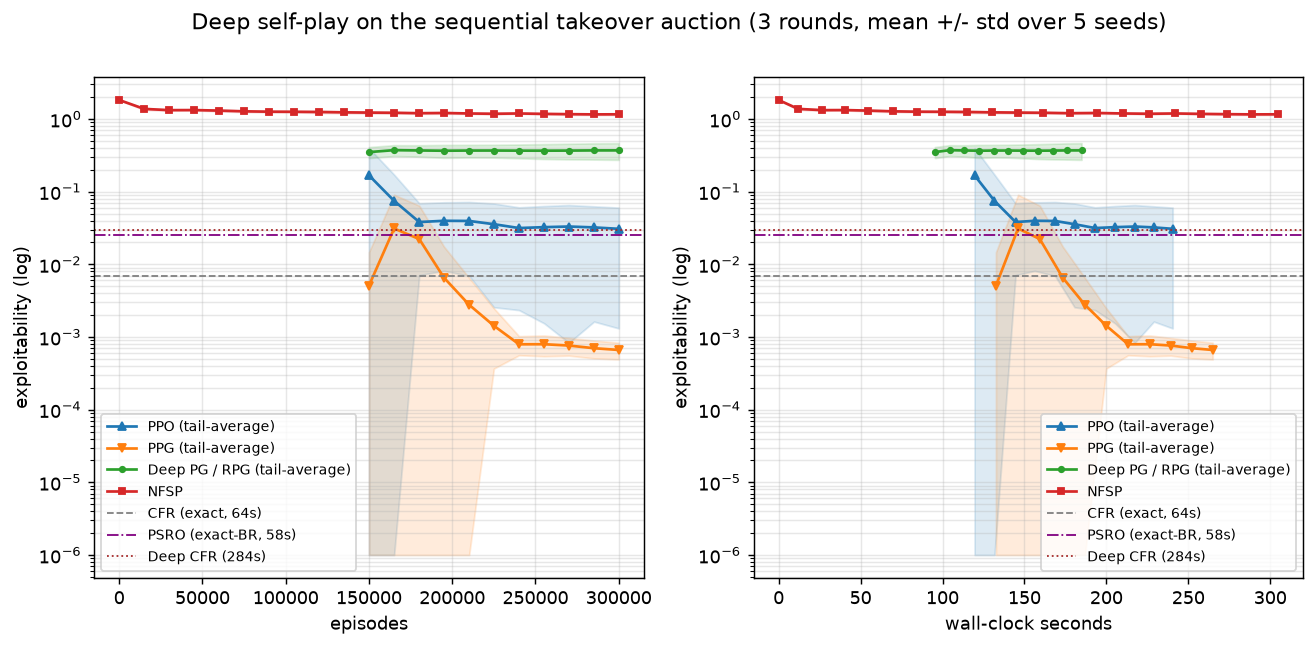}
\caption{Deep self-play on the three-round takeover auction (mean $\pm$ std over five seeds). PPG converges to a near-exact equilibrium ($0.0007$ exploitability), below the exact-CFR reference at the budget shown; PPO is competitive but seed-sensitive; NFSP does not converge within the shared budget.}
\label{fig:seqdeep}
\end{figure}

On a three-round instance small enough that exact counterfactual regret minimization and exact exploitability remain a valid reference ($\mathrm{num\_values}=3$, $\mathrm{num\_bids}=6$, about $6{,}655$ states), we run deep self-play for PPO, PPG, and deep policy gradient, alongside NFSP, and compare against CFR, PSRO, and Deep CFR, scoring each by exact exploitability of its tabular tail-average over five seeds (Figure~\ref{fig:seqdeep}).

The ordering is stark. PPG reaches $0.0007\pm0.0002$ exploitability, with all five seeds in agreement, dipping below the exact-CFR reference of $0.0069$ at the budgets shown (we do not read this as PPG beating CFR at equilibrium-finding: CFR run longer would go lower, and it reached its value faster; rather, PPG converges to a genuine near-exact equilibrium and is the strongest deep method by a wide margin). PPO is competitive at $0.031$ but seed-sensitive ($\pm0.029$). Deep policy gradient plateaus near $0.37$. NFSP, a standard imperfect-information baseline, does not converge within the shared $300{,}000$-episode budget, crawling from $1.83$ only to $1.16$; it was already the weakest method on our sealed game, and the sequential structure amplifies the gap. Absolute exploitability is not comparable across the sealed and sequential games; only the within-game ordering is.

\begin{figure}[H]
\centering
\includegraphics[width=\textwidth]{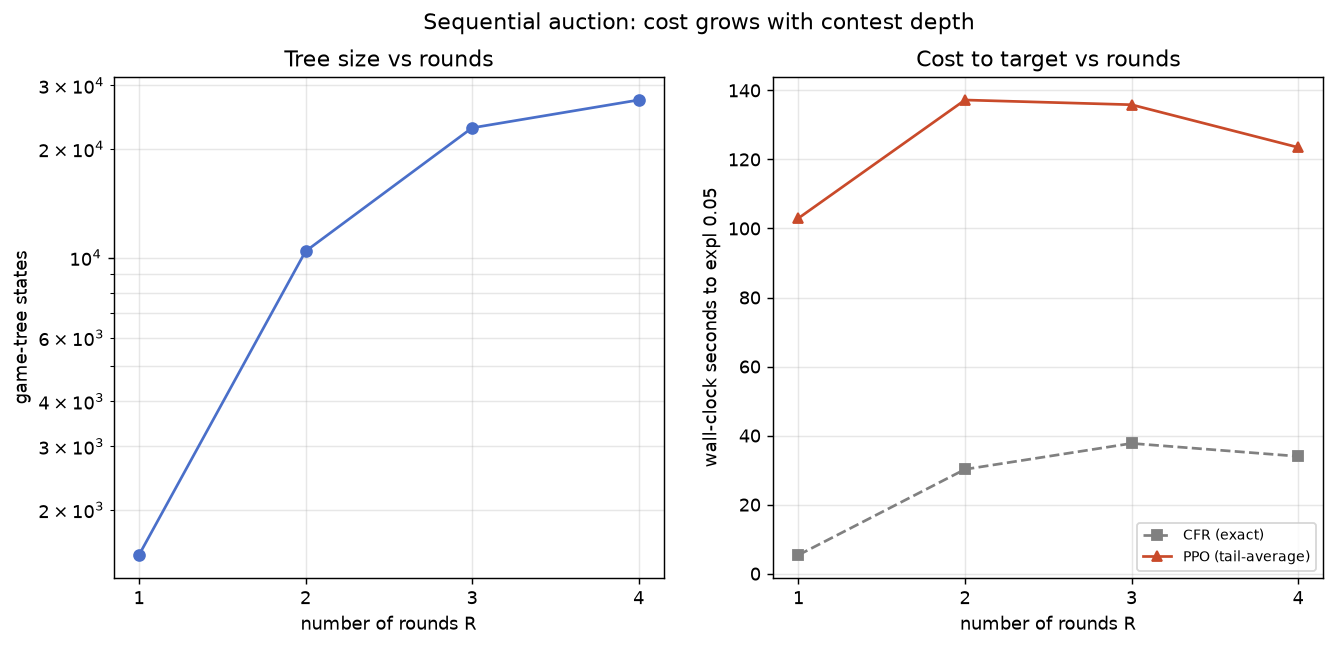}
\caption{Scaling in the number of rounds. Left: game-tree size grows steeply with rounds (the flattening at $R=4$ is the price-grid ceiling beginning to bind). Right: exact CFR's wall-clock cost to reach exploitability $0.05$ jumps from one round to two and then plateaus, while PPO's is roughly depth-invariant.}
\label{fig:seqscaling}
\end{figure}

\paragraph{Scaling in the number of rounds.} Because the ascending-bid ceiling caps tree depth at the price grid, we widen the grid ($\mathrm{num\_bids}=8$) so that added rounds genuinely deepen the tree, and measure the wall-clock each solver needs to reach an exploitability of $0.05$ as the contest grows from one to four rounds (Figure~\ref{fig:seqscaling}). The tree grows about eighteen-fold over the range. Exact CFR remains faster than PPO at every round count in this tractable range, but its cost climbs steeply from one round to two and then plateaus ($5.4$, $30.2$, $37.7$, $34.0$ seconds) while PPO's stays roughly flat ($103$, $137$, $136$, $124$ seconds), so the ratio of CFR to PPO wall-clock narrows from $0.05$ at one round to $0.27$ at four: the gap closes but does not cross. The crossover belongs to the regime where the game can no longer be enumerated at all.

\section{The intractable regime}
\label{sec:intractable}

\begin{figure}[H]
\centering
\begin{minipage}[t]{0.48\textwidth}
\centering
\includegraphics[width=\textwidth]{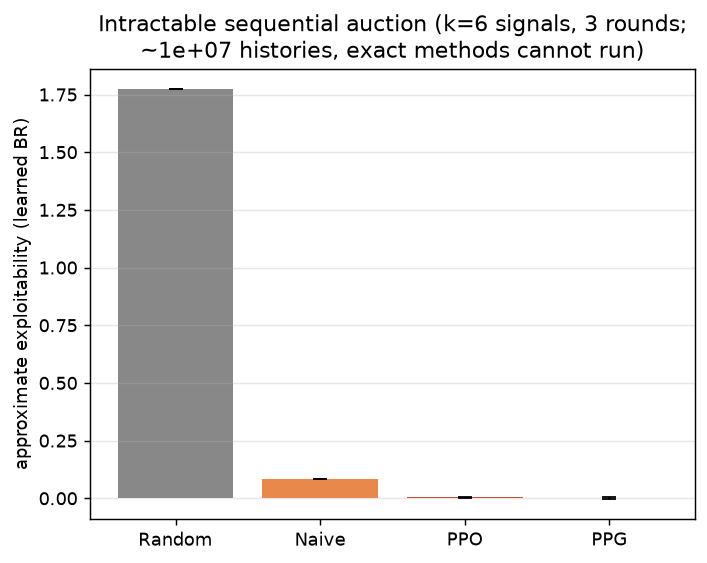}
\captionof{figure}{Intractable multi-round, multi-signal auction ($k=6$ signals, three rounds, about $10^{7}$ histories). PPO ($0.004\pm0.003$) and PPG ($0.003\pm0.004$) drive the learned-best-response exploitability estimate below a naive unshaded bidder ($0.083$) and far below uniform play ($1.77$), in a regime where exact solvers cannot run.}
\label{fig:seqintr}
\end{minipage}\hfill
\begin{minipage}[t]{0.48\textwidth}
\centering
\includegraphics[width=\textwidth]{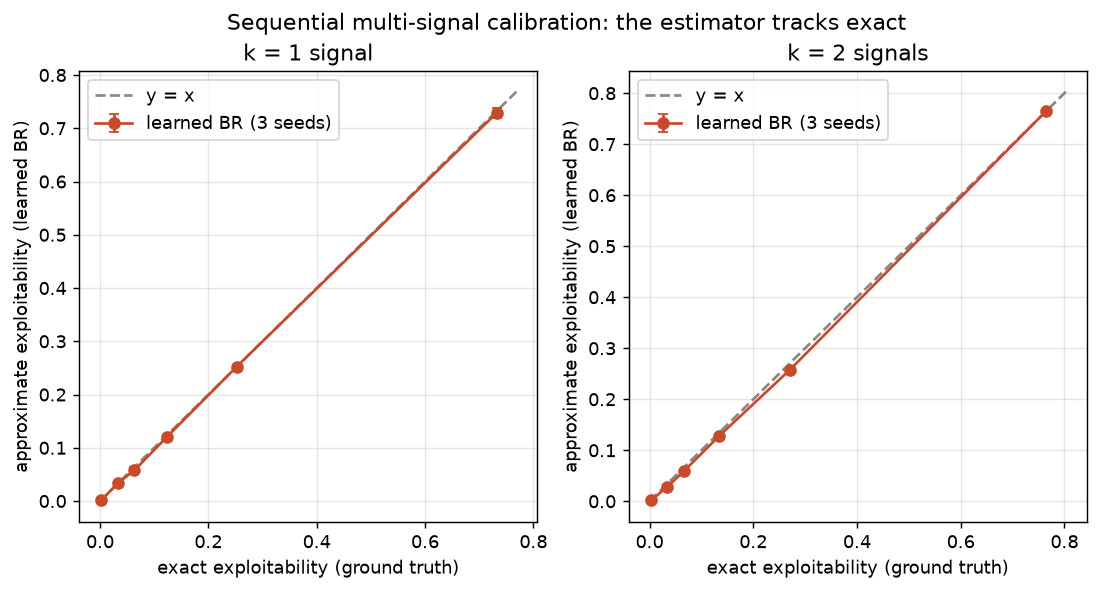}
\captionof{figure}{Estimator calibration on enumerable multi-signal sequential games ($k=1,2$). The learned-best-response estimate tracks exact exploitability across policies of known exploitability (CFR equilibrium mixed with uniform play), lying close to the $y=x$ line over the whole range.}
\label{fig:seqcalib}
\end{minipage}
\end{figure}

Everything so far has been small enough to write down completely, which is what makes the exact certificates possible. Real contests are not, so the last computational question is whether any of this survives once the game is too big to enumerate. Adding rounds and adding private signals both enlarge the game, but only the signals do so without bound: information sets grow as $\mathrm{num\_values}^{\,k}$ in the signal count $k$, and the history count as $\mathrm{num\_values}^{\,1+2k}$ times the bidding subtree, so a multi-round, multi-signal instance quickly outruns exact enumeration. Giving each bidder six pieces of private due-diligence information instead of one is enough to do it. On an instance with $k=6$ signals per bidder over three rounds (on the order of $10^{7}$ histories, beyond exact CFR and exact exploitability), we train PPO and PPG self-play without tabulating and score them with a learned-best-response exploitability estimate: freeze the policy, train a PPO best-responder against it, and Monte-Carlo the gain from deviating. A learned best response underestimates the true one, so the estimate is a lower bound, which we report as such rather than as a Nash certificate. Against this estimator, uniform-random play is highly exploitable ($1.77$) and a naive no-shading bidder economically so ($0.083$), while PPO and PPG drive the estimate to $0.004\pm0.003$ and $0.003\pm0.004$ over three seeds, below the naive bidder and down to the estimator's calibrated resolution floor, consistent with near-unexploitable play, in a regime where the exact solvers cannot run (Figure~\ref{fig:seqintr}). The two learners are not separable from each other at that spread; both are simply at the floor.

To show the estimate is trustworthy on the sequential multi-signal information-set structure, and not merely reading a weak best response, we calibrate it on enumerable instances ($k=1,2$) where exact exploitability is available as ground truth: we build policies of known exploitability by mixing the CFR equilibrium with uniform play and confirm the learned estimate tracks the exact value (Figure~\ref{fig:seqcalib}). The tracking is tight across the whole range: at $k=2$, for example, the estimate reads $0.257$ against an exact $0.270$ at a $40\%$ uniform mix and $0.765$ against an exact $0.764$ at full uniform, and at the CFR equilibrium both fall to the estimator's resolution floor of a few thousandths. The estimate is therefore a lower bound calibrated on the sequential multi-signal structure at $k=1,2$, not an artifact of a weak best-responder. It is applied at $k=6$, where no exact ground truth exists to calibrate against; the calibration is evidence that the estimator tracks the truth on this game's information-set structure, not a guarantee at a signal count we cannot enumerate.

\section{Discussion and limitations}
\label{sec:limitations}

The economic headline is a cautionary one for practice and theory alike: the toehold's \emph{effect on} deterrence, the thing that would justify acquiring one, is a feature of our model truncated at a single round and does not survive a genuine second round. Deterrence itself stays high at two rounds; what it stops doing is responding to the toehold. by the third the equilibria multiply and it is not identified at all. This does not say toeholds are worthless (they still raise the holder's profit), but it removes the identified strategic edge that the puzzle literature has struggled to reconcile with their rarity.

Put concretely, the two reasons usually given for taking a toehold do not fare alike here. The profit reason holds up: in every configuration we solve, a larger toehold leaves its holder better off, and that is the robust part of the classical story. The deterrence reason does not. In our model a toehold does not reliably make a rival quit, because whether the rival quits turns on whether it reads an aggressive opening as strength, and that reading is a matter of which equilibrium the two sides fall into rather than something the toehold causes. The same auction supports a rival who folds and a rival who does not, at the same toehold, with the holder earning the same either way. So a deal team that is pricing the disclosure and market-impact costs of a toehold against an expected deterrence benefit is pricing them against a benefit our model cannot deliver. The point is about attribution rather than absence. Rivals do fold in this auction, often at a third of the time, and at our two largest toeholds every certified equilibrium deters. But the model never credits the toehold for it, because the same folding appears at a toehold of zero, and where deterrence is pinned down it is pinned at the value the toehold-free contest already delivers. We are describing a model, not a market, and the limitations below are blunt about the distance between the two. But the direction of the correction is clear, and it happens to point the same way the data does: toeholds are rare, and one reason may simply be that the edge they are supposed to confer was never as solid as the short models suggested.

Several limitations bound the claims, and they are worth being blunt about.

\emph{One instance.} Every economic result above is one parameterization: three value levels, a nine-level price grid, one signal per bidder, and symmetric signal noise of $0.5$. We sweep the toehold, the number of rounds, and the grid resolution, and nothing else. We have not varied the number of values, the number of signals, or the noise, so we do not know whether the multiplicity or the two-move saturation survives a richer signal space. Where we write that the deterrence channel is an artifact of stopping at one or two moves, that is a statement about \emph{our} model truncated at one or two moves, not a theorem about the class of short-contest models; we solve neither Dodonova's model nor Bulow-Huang-Klemperer's, and reproducing their comparative static at $R=1$ is not the same as deriving their result.

\emph{Per-configuration, not theorems.} The value-is-pinned observation is verified where measured and explicitly not a theorem: it fails at one and two rounds with a zero toehold, where multiple values coexist. Nor is the non-identification an econometric claim. What we exhibit is a set of computed equilibria at one instance that a single deterrence number cannot summarize; that is model-side non-identification, and it does not by itself establish that deterrence is unidentifiable in field data.

\emph{Precision.} The $\varepsilon$-equilibria we exhibit are certified to $\varepsilon$ between $5\times10^{-7}$ and $8\times10^{-5}$, small but not exact; in a flat game an $\varepsilon$-equilibrium can sit some distance from an exact one, so we phrase every claim in terms of $\varepsilon$-equilibria that a referee can reproduce. The intractable-regime numbers are learned-best-response \emph{lower bounds}, calibrated at $k=1,2$ and applied at $k=6$, not certificates.

\emph{Grid robustness is partial.} The refinement study establishes that the deterrence multiplicity is not a coarse-grid artifact, since it appears at nine levels and persists at thirteen. It does not establish it at a positive toehold under refinement. Six restarts certify nothing at thirteen levels at either $\theold=0.15$ or $\theold=0.30$. Because a cell that certifies nothing may be a hard cell or merely an unsampled one, we re-ran $\theold=0.15$ at thirteen levels with twelve additional starting policies, for eighteen solves in total at the certified budget, twelve of them new. Exactly one clears the bound, at deterrence $0.333$ and $\varepsilon=3.8\times10^{-5}$. So the fine grid does yield a certified equilibrium at a positive toehold, but it yields one conduct rather than two, and the only refined cell that exhibits the multiplicity itself remains the zero-toehold one, where two of six restarts are certified. The extra restarts do reach low-deterrence conduct repeatedly, four of the twelve settling below $0.013$, but none of those solves comes close to the bound: the tightest is $2.7\times10^{-3}$, twenty-seven times the acceptance threshold, and two of them are still improving when the budget stops, their tightest iterate being their last, so we cannot say whether they are slow or non-convergent. A reader should therefore take the refined claim as holding where we can certify it and as untested, rather than tested and passed, at positive toeholds on the fine grid. The finer cells are the slowest solves in the suite, at roughly a hundred hours each on one core by our own measurement, which is why this study is bounded by compute rather than by method. What the extra restarts change is the diagnosis: the obstacle at $\theold=0.15$ was at least partly sampling rather than budget, since the profile that certifies was found by a new starting policy at the same budget six earlier ones failed at. The headline does not rest on this study, which rests instead on the rounds sweep at $\varepsilon\sim10^{-6}$.

\emph{The premium intervals are lower bounds.} The spread we report for the jump premium at each toehold is the range across the certified profiles that six restarts happen to find. It is not the range across the equilibrium set, which we do not enumerate and cannot bound: a seventh restart could widen any of these intervals, and none of them can narrow. This cuts in the direction of our claim rather than against it, since the claim is that the premium is not pinned down, but it does mean the endpoints should be read as observed extremes and not as the support of anything.

\section{Conclusion}

We solved a multi-round ascending takeover auction with a toehold and found that the toehold's supposed deterrence edge is not what three decades of two-move intuition suggests. Equilibria are multiple and give the same profit with different conduct; preemption occurs even without a toehold; and the clean toehold-to-deterrence relationship is an artifact of stopping at two moves, dissolving as soon as the contest has a genuine second round and leaving deterrence unidentified by the third. Preemptive jump bidding is real and computationally reproducible, but as Fishman-style signalling, not as a toehold effect, which offers a model-side contribution to the toehold puzzle rather than a resolution of it. Methodologically, forced-opening best responses read the economics without trusting any single solver, and a fictitious-play solve taken at face value manufactures confident artifacts. On the computational side, a generic policy-gradient method solves the sequential game where a standard imperfect-information baseline does not, and the analysis extends into a regime beyond exact enumeration. We release the games, solvers, and experiments so the results can be reproduced and extended.

\bibliographystyle{plainnat}


\end{document}